\documentclass[reprint,cha,groupedaddress]{revtex4-1}

\usepackage{graphicx}
\usepackage{amssymb}
\usepackage{placeins}
\usepackage[T1]{fontenc}
\usepackage{color}
\usepackage{url}
\usepackage{rotating}
\usepackage[squaren]{SIunits}
\usepackage{amsmath}
\usepackage{wasysym}
\usepackage[latin1]{inputenc}
\usepackage[T1]{fontenc}
\usepackage{ae,aecompl}
\usepackage{xcolor}
\newcommand\TI{\mathit{TI}}
\newcommand\cc{\mathit{cc}}

\begin{document}

\title{Stability of Synchrony against Local Intermittent Fluctuations in Tree-like Power Grids}
\author{Sabine Auer}
 \email{auer@pik-potsdam.de}
 \affiliation{Potsdam Institute for Climate Impact Research, 14412 Potsdam, Germany}
 \affiliation{Department of Physics, Humboldt University Berlin, 12489 Berlin, Germany}
\author{Frank Hellmann}%
\affiliation{Potsdam Institute for Climate Impact Research, 14412 Potsdam, Germany}%
\author{Marie Krause}
\affiliation{Technical University of Berlin, Institute of Mathematics, 10587 Berlin}%
\author{J\"urgen Kurths}
\affiliation{Potsdam Institute for Climate Impact Research, 14412 Potsdam, Germany}
\affiliation{Department of Physics, Humboldt University Berlin, 12489 Berlin, Germany}
\affiliation{Department of Control Theory, Nizhny Novgorod State University, 606950 Nizhny Novgorod, Russia}
\affiliation{Institute of Complex Systems and Mathematical Biology, University of Aberdeen, Aberdeen AB24 3FX, UK}

\begin{abstract}
80\% of all Renewable Energy Power in Germany is installed in tree-like distribution grids. Intermittent power fluctuations from such sources introduce new dynamics into the lower grid layers. At the same time, distributed resources will have to contribute to stabilize the grid against these fluctuations in the future. In this paper, we model a system of distributed resources as oscillators on a tree-like, lossy power grid and its ability to withstand desynchronization from localized intermittent renewable infeed.

We find a remarkable interplay of network structure and the position of the node at which the fluctuations are fed in. An important precondition for our findings is the presence of losses in distribution grids. Then, the most network central node splits the network into branches with different influence on network stability.
Troublemakers, i.e. nodes at which fluctuations are especially exciting the grid, tend to be downstream branches with high net power outflow. 
For low coupling strength, we also find branches of nodes vulnerable to fluctuations anywhere in the network. These network regions can be predicted at high confidence using an eigenvector based network measure taking the turbulent nature of perturbations into account. 

While we focus here on tree-like networks, the observed effects also appear, albeit less pronounced, for weakly meshed grids. On the other hand the observed effects disappear for lossless powergrids often studied in the complex systems literature. 
\end{abstract}


\pacs{Valid PACS appear here}

\maketitle

\begin{quotation}
The full decarbonization of the energy sector by 2050 is non-negotiable to meet the emission targets of the Paris agreement \cite{rogelj2016paris}. Hence, the effort to deploy Renewable Energy Sources in the electricity, transport and heat sector continues and soon, the power system will undergo a regime shift from central conventional to distributed power production. For power grid operators this means: instead of centrally controlling and distributing large amounts of power from few power plants to the lower grid levels, the new challenge is to control lots of small generation units in a swarm-type manner. So far, the understanding of the novel dynamics of intermittent power resources in these lower grid levels, called distribution grids, is poorly understood because so far, there was no need to. However, the rapid approach of the regime shift asks for actions. This paper focuses on the stability of synchrony in distribution grids against local intermittent fluctuations at single nodes in the network and identifies network regions that are especially susceptible or infectious towards power fluctuations. Such insights will help to develop control techniques that are feasible and cost-efficient at the same time.
 \end{quotation}

\section{Introduction}
The increasing share of Renewable Energy Sources (RES) poses a wide range of challenges for power grid stability. Today, in Germany 80\% of all installed power from RES lies in distribution grids. With a growing number of especially wind and solar power plants, new dynamics are introduced into the lower grid layers which need to be understood.
In the following, we investigate the influence of the power grid's topology and the placement of variable renewable infeed on stochastic stability measures in distribution grids. Our focus will be the stability of the synchronous state \cite{arenas2008synchronization,dorfler2014synchronization} and the ability of the system to keep frequency fluctuations small.
\par

Work on the effect of stochastic fluctuations from RES on grid stability has been started very recently and thus, only few publications on this matter exist. 
For lossless power grids, the recent work \cite{Xiaozhu2017} studied analytically the influence of single-node monochromatic oscillations and its spreading throughout the network. There, three frequency regimes were identified: a bulk, resonant and local regime. In the bulk regime for low frequencies the network is excited as a whole, whereas in contrast, high frequency perturbations stay localized at the fluctuating node and decay exponentially with network distance. Most interestingly, for the mid-frequency region the network topology was identified to play a major role in network stability as complex, non-trivial patterns of stability and instability emerge.
This analytic work is an important basis for understanding the influence of intermittent noise on power grid stability where fluctuations cover the whole frequency spectrum with a Kolmogorov-like turbulent power spectrum \cite{milan2013turbulent,woyte2007fluctuations,anvari2016short}. \par In \cite{schmietendorf2016} the impact of the detailed stochastic properties of the RES infeed were studied in great detail. There, the authors compared white Gaussian noise, Gaussian noise with a turbulent power spectrum and intermittent noise. The latter noise displays a fat tail in both the power spectrum and the increment time series, which corresponds to a (long-range) correlation in time. The work showed that the time-correlated noise leads to the strongest network destabilization. Noise of the same power spectrum but without intermittency induces smaller frequency deviations. Thus, it is most important to include the intermittent nature of RES power fluctuations in future grid stability analysis.
\cite{Xiaozhu2017,schmietendorf2016} already mentioned the importance of the topology aspect. \cite{Matthiae2016} showed for two nodes (with white-noise power fluctuations) coupled to a bulk grid how the node with lower coupling strength is the one destabilizing the network. Moreover, with Kramer's escape rate theory the authors demonstrated how in a network with all machines subject to white noise, the weakest links, identified by the ``saddles'' of the grid, tend to be the overloaded lines (links with phase difference exceeding $\pi/2$). Thus, in this study the stability focus was on heavily-loaded lines.

In contrast to previous studies, here we specifically model distribution grids (see Section \ref{sec:model}) with intermittent fluctuations localized at a single node and a focus on how the position of the perturbed node in the network influences its ability to corrupt the whole power grid system.

In this way we can classify branches of nodes as troublemakers and fluctuation sensitive nodes. The net power outflow of each branch determines to what extent power fluctuations at associated nodes tend to cause notable frequency fluctuations at all network nodes. We call these nodes troublemakers or drivers of instability. The other classification describes nodes that react with notable frequency fluctuations irrespective of which node introduces the power fluctuations into the network. Such nodes appear especially pronounced for low coupling strength and can be identified with a novel network measure based on a turbulent weighting of the eigenvectors of the Jacobian. \par

In the following, we consider the model cases of a microgrid described in Section \ref{sec:ModelCases} together with a description of the intermittent noise in \ref{sec:noise}. In Section \ref{sec:stab_measures} we discuss adequate measures for stability in such a stochastic modeling setup. Finally, Section \ref{sec:results} presents the results which are discussed in Section \ref{sec:discussion}.

\section{Model Setup} \label{sec:model}
\subsection{Distribution Grid Model}\label{sec:dist_grid_model}

Our model is designed to represent distribution grids. Thus, we chose tree-shaped networks as the underlying topology (generated with a random growth model \cite{schultz2014random}) and introduce lossy lines, since the common assumption of non-lossy lines for transmission grids does not hold for distribution grids.

For the power grid dynamics, we assume a grid which is dominated by inverters because we want to analyze a scenario with high RES penetration and wind and solar power plants that are connected to the grid via inverters. The classical power grid model (or swing equation) is derived from the Synchronous Machine Model representing conventional generators and their rotating masses \cite{nishikawa2015comparative}. Inverters and their power electronics may be programmed as Virtual Synchronous Machines by using a smooth droop control. This then leads to the same equations for the voltage angle $\phi$ and frequency $\omega$ in terms of the (virtual) inertia $H$, power infeed $P$, (virtual) damping $\alpha$, line susceptabilities, $Y=G+jB$, and voltage magnitudes $U$ \cite{schiffer2013synchronization}:
\begin{align}
\begin{split}
\dot{\phi}_i=&\omega_i,\\
\dot{\omega_i}=&\frac{1}{H}(P_i+\Delta P(t)\delta_{ik}-\alpha\omega_i\\
 &-\sum_jU_i|Y_{ij}|U_j \sin(\phi_i-\phi_j+\phi_{ij})).
\end{split}
\label{eq:dynmics}
\end{align}
The virtual inertia and damping for the network model is given by the low-pass filter exponent $\tau_p$ and the droop control parameter $k_p$ from grid-forming inverters: $H = \tau_p/k_p$, $\alpha=1/k_p$, $\forall i$ with $i=1,..,N$. The power fluctuations $\Delta P(t)$ is only applied at node $k$.
The impedance of the lines for typical Mid-Voltage grid lines with 20 kV base voltage equals $Z= Y^{-1}= (0.4 + 0.3j) \Omega/km$. The coupling strength between a node pair $(i,j)$ then equals $U_i|Y_{ij}|U_j$. The addition of the resistance leads to line losses and at the same time introduces a phase shift of $\phi_{ij}\approx \arctan(\frac{G_{ij}}{B_{ij}})$ that does not occur in the non-lossy model but will be shown to have significant consequences for stability. The Jacobian of \eqref{eq:dynmics} equals
\begin{equation}
J_{ij}=
\begin{bmatrix}
\frac{\partial \dot{\phi_i}}{\partial\phi_j}&\frac{\partial \dot{\phi_i}}{\partial\omega_j}\\
\frac{\partial \dot{\omega_i}}{\partial\phi_j}&
\frac{\partial \dot{\omega_i}}{\partial\omega_j}
\end{bmatrix}.
\label{eq:jac}
\end{equation}

For each simulation run, the same intermittent time series (the sum of wind and solar power fluctuations, scaled with the power production) was added to a single node's power input.  

The equation is written in a co-rotating frame, thus the synchronous state we study is characterized by $\omega_i = 0$. Hence, we are interested in the deviations from the stable frequency set point which corresponds to $50$Hz. In the following, we mainly investigate $\Delta f = \omega/(2\pi)$

\subsection{Microgrids as a Model Case}\label{sec:ModelCases}

In order to understand the abillity of a distributed system to maintain synchrony, we will study the case of an islanded microgrid that is internally balanced and not connected to a higher grid level. Islanded microgrids play a role for the decentral provision of energy, but also as part of a safety and stability strategy to localize faults by partitioning the grid into autonomous units.

A microgrid has a non-hierarchical grid topology. The grid is balanced within itself, in our case there are 50 producers and  50 consumers with $P_i=\pm 0.2 MW$ power infeed before losses. The power infeeds are chosen homogeneously to isolate topology and network effects in the model. As there is no connection to upper grid levels, losses are compensated locally at each node, and the net power infeed is given by $\tilde{P}_i = (P_i +P_{loss}/N)$.
Each node has $H_i=0.099s^2$.

\begin{figure}[b!]
\centering
\includegraphics[width=0.23\textwidth]{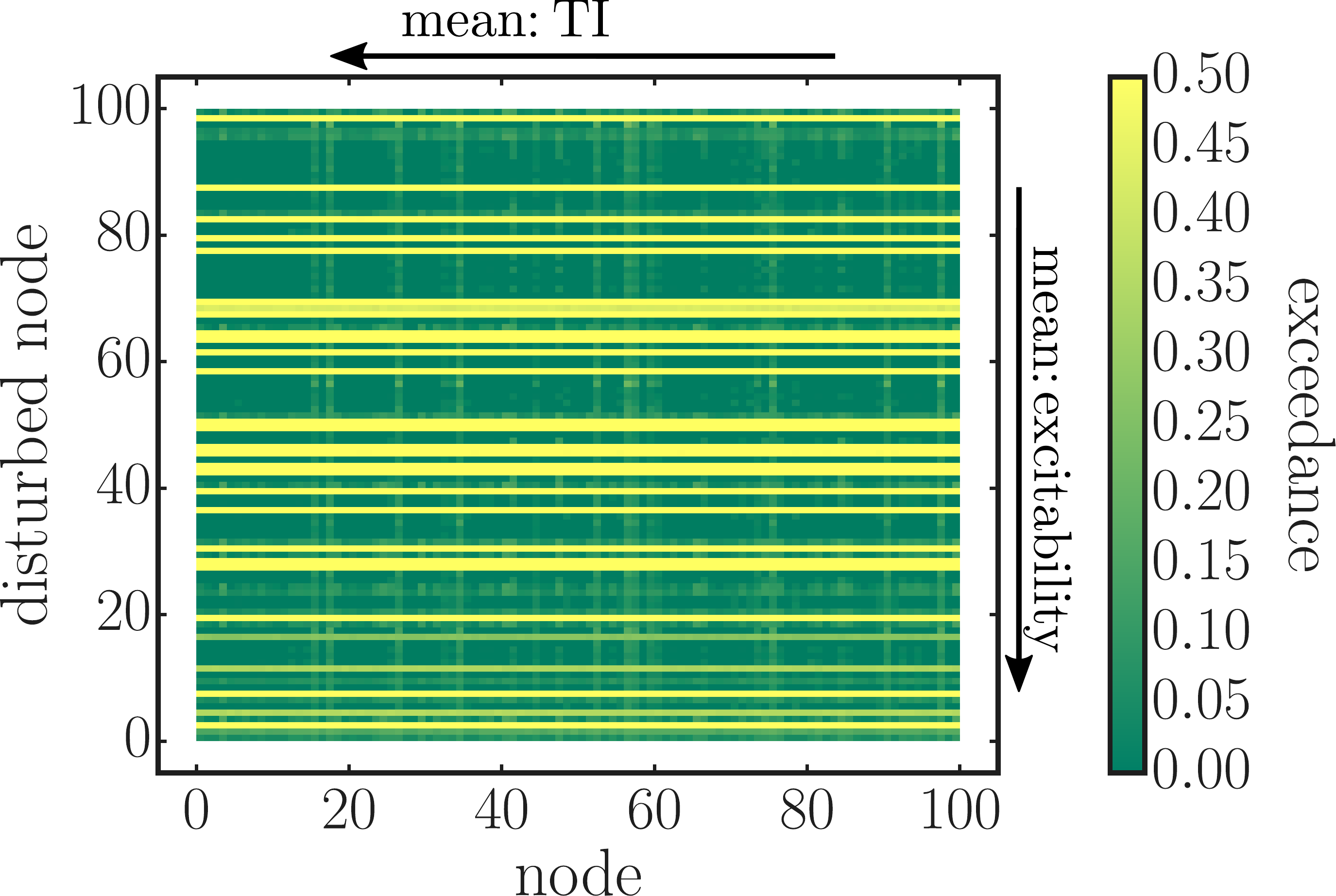}
\hspace{0.01\textwidth}
\includegraphics[width=0.225\textwidth]{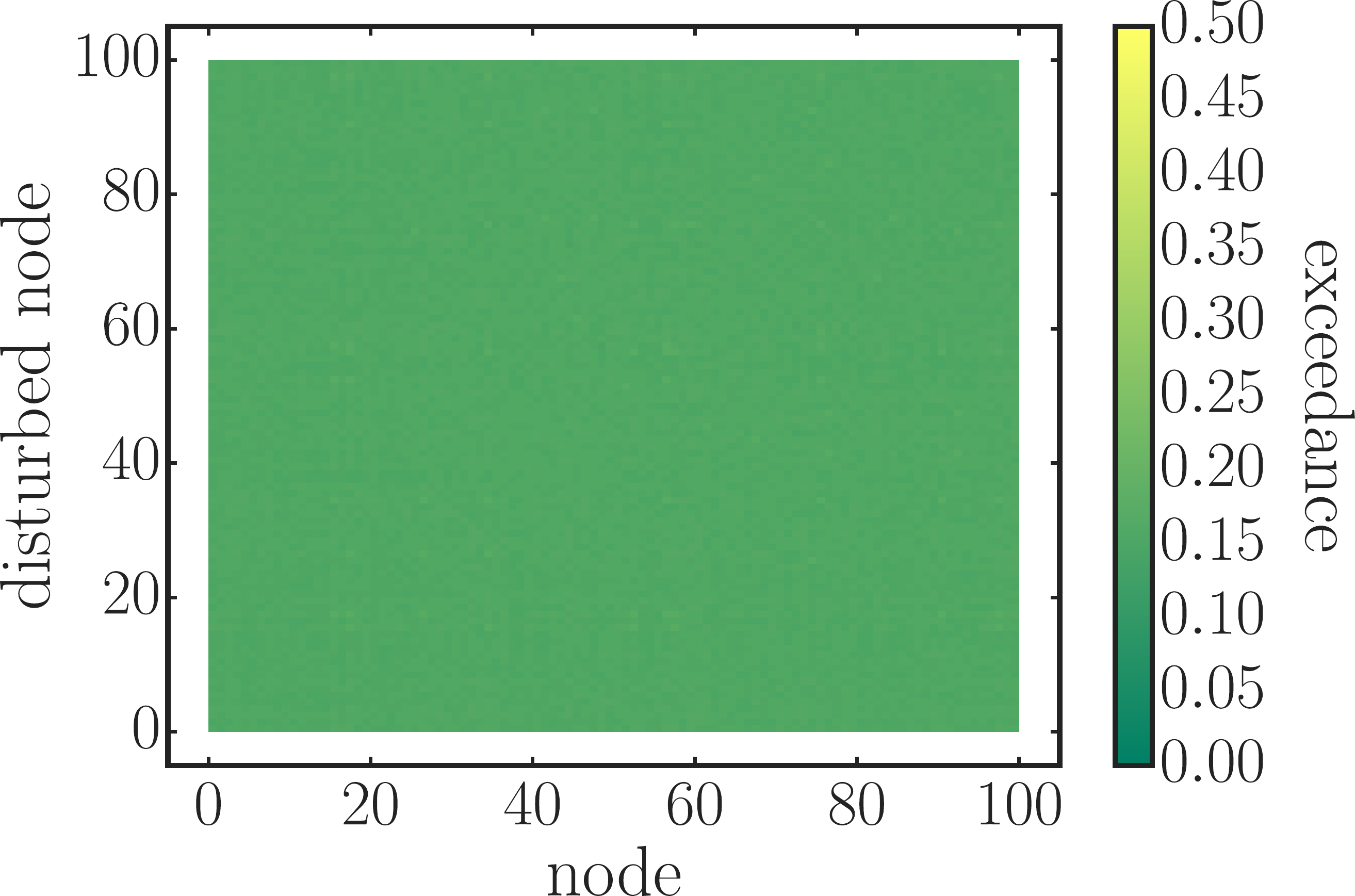}
\caption{Color Plot of Single Node Exceedances (see \eqref{eq:exc}) for each simulation run or disturbed node (y-axis) and each node in the network (x-axis) with (left) lossy and (right) non-lossy lines: $Z=Y^{-1}=(0.4+0.3j) \Omega/km$ and $Z=(0+0.3j) \Omega/km$, respectively}
\label{fig:stab_measure}
\end{figure}

\subsection{Intermittent Noise}
\label{sec:noise}

In the following simulations the intermittent time series for solar and wind power fluctuations were generated by a clear sky index model, based on a combination of a Langevin and a Jump process, developed in \cite{anvari2016short}, and a Non-Markovian Langevin type model developed in \cite{schmietendorf2016}, respectively. An example time series, $\Delta P(t)$, of the combined wind and solar power fluctuations, $\Delta P_W(t)$ and $\Delta P_S(t)$ respectively, is shown in Fig.\ref{fig:micro_net_exc} (left)
\begin{equation}
\Delta P(t) = 0.5 \Delta P_W(t)+0.5 \Delta P_S(t).
\end{equation}

The stochastic nature of such processes were identified in \cite{anvari2016short, schmietendorf2016} with the help of time series analysis. Important characteristics are the probability distribution function (PDF), the increment distribution and the power spectrum. 
If the PDFs, of both the time series of power and power increments, are fat tailed (the tails are not exponentially bounded \cite{asmussen2008applied}), we define this as intermittency.
Also, the power generation from wind and solar power plants has a power spectrum that is power-lawed with the Kolmogorov exponent of turbulence \cite{milan2013turbulent,anvari2016short}. Thus, time series from such sources show long-term temporal correlations. The Hurst exponent $0 < H_q < 1.0$ is a time-lag dependent measure for such long-time memory in a time series \cite{hurst1956problem,alessio2002second,preis2009accelerated,carbone2004analysis}. It quantifies the rate at which the autocorrelations of the time series decreases with increasing time lag. In this work, we display the turbulent and intermittent nature of RES the with long-term positive autocorrelation $H_q$ close to unity and fat-tailed increment distributions.

\subsection{Stochastic Stability Measures}\label{sec:stab_measures}

\begin{figure*}[t!]
\begin{minipage}{0.7\textwidth}
\centering
\includegraphics[width=0.47\textwidth]{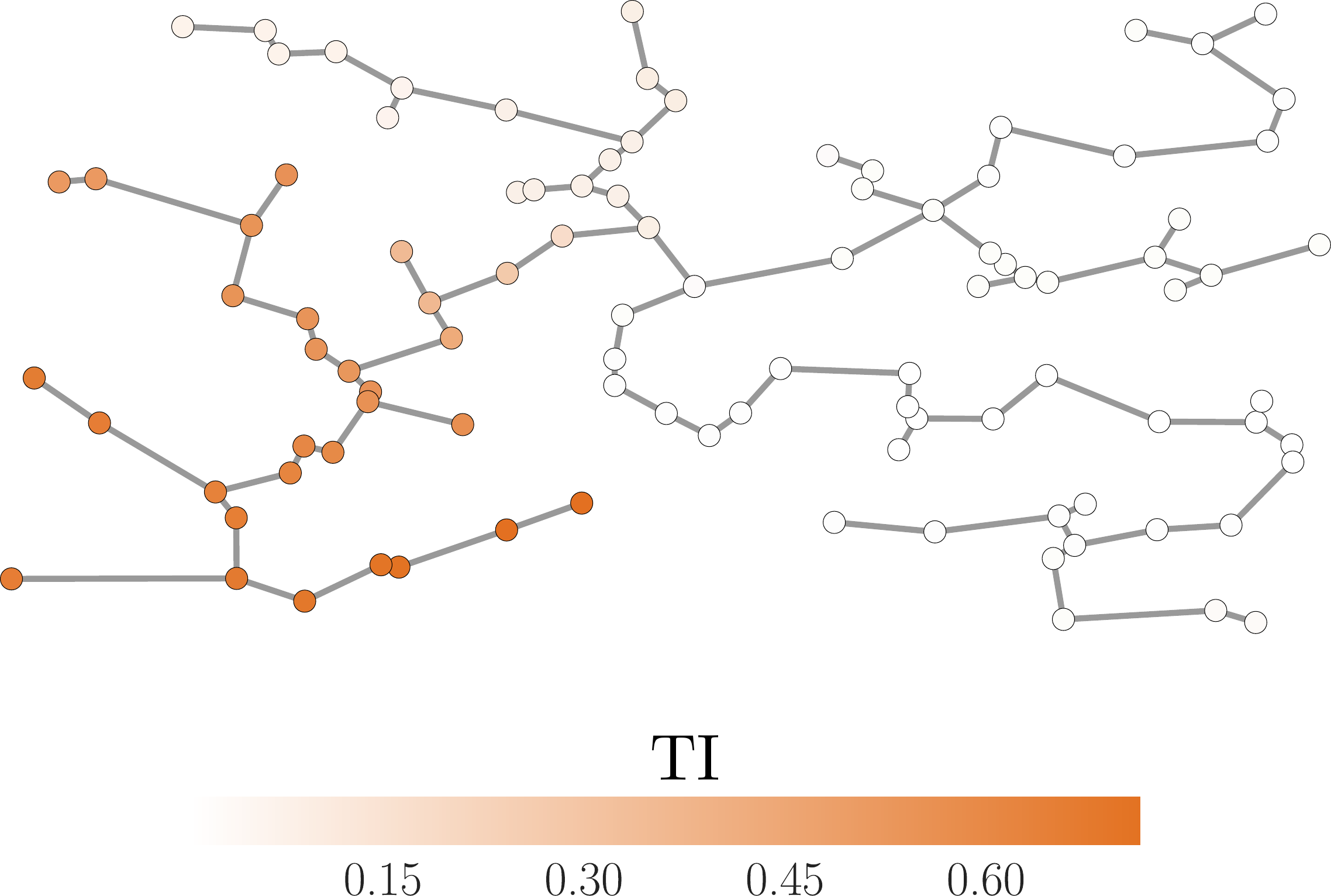}\hspace{0.05\textwidth}\includegraphics[width=0.47\textwidth]{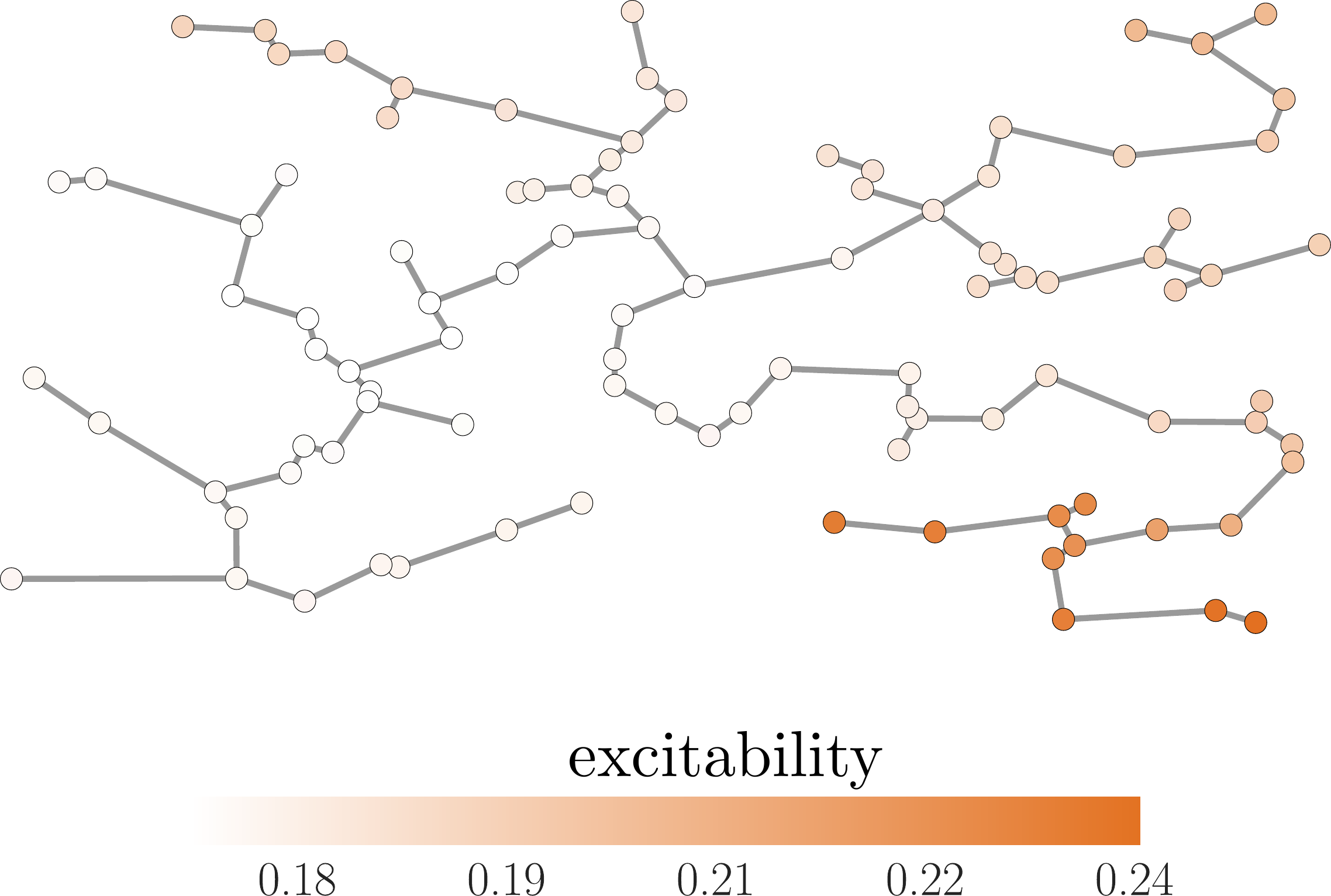}
\vspace{0.05\textwidth}\\\includegraphics[width=0.47\textwidth]{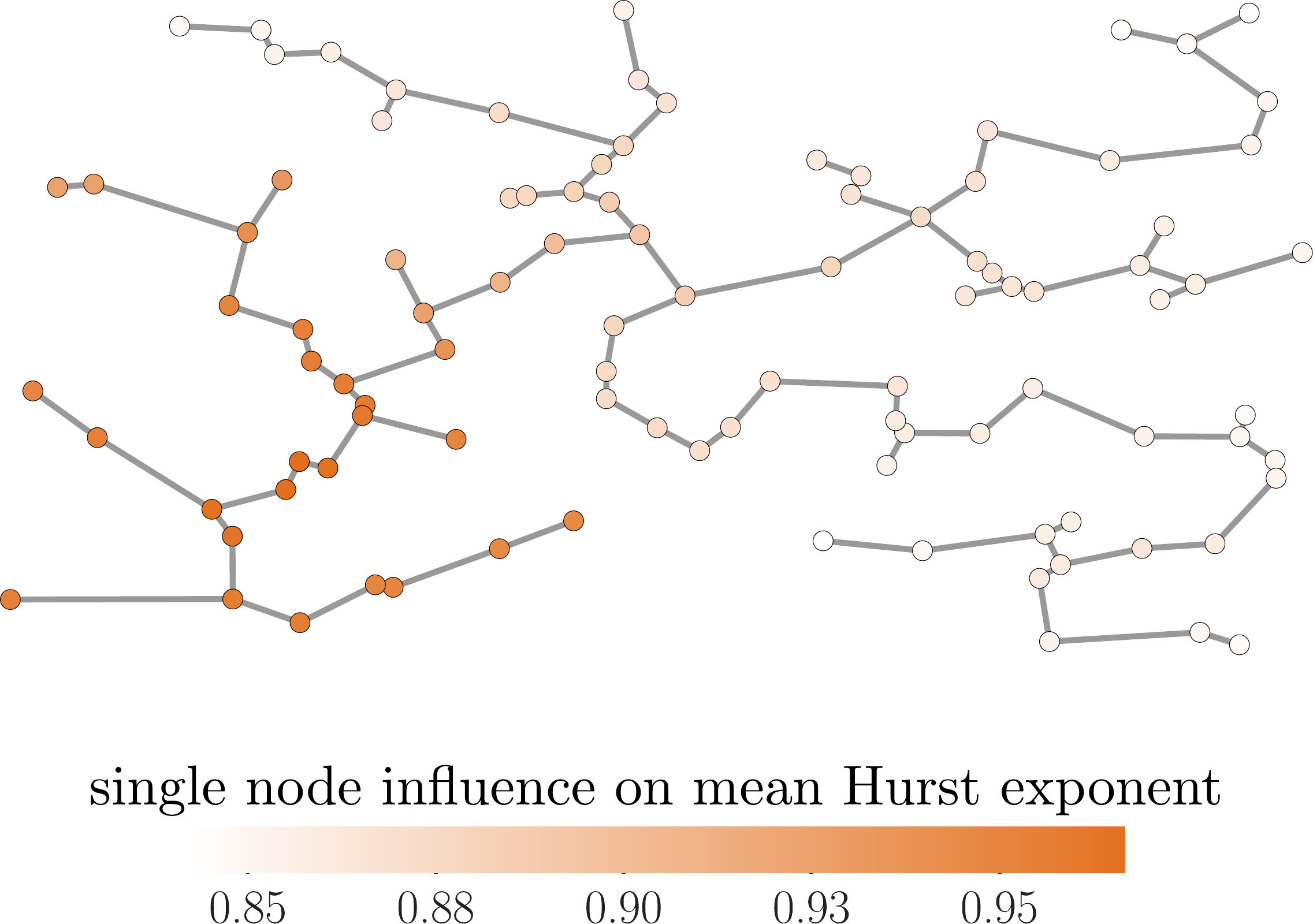}\hspace{0.05\textwidth}\includegraphics[width=0.47\textwidth]{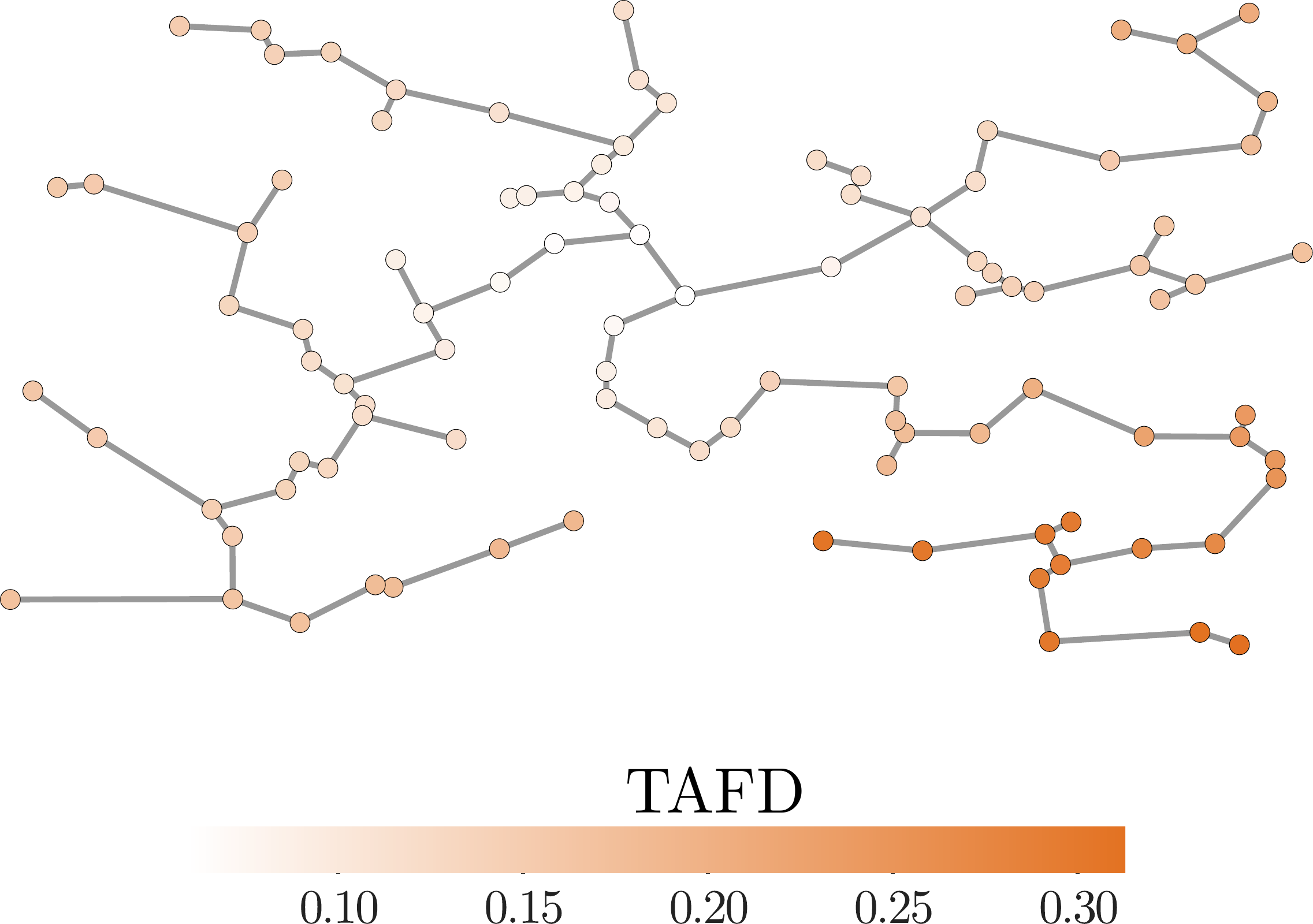}\end{minipage}
\hspace{0.04\textwidth}
\begin{minipage}{0.22\textwidth}
\caption{Top left: The Troublemaker Index ($\TI$, see\eqref{eq:ti}) as the mean over the x-axis of the colorplot. Top right: Excitability (see \eqref{eq:excitability}) is the mean over the y-axis of the colorplot.
Bottom left: Network plot of single node influence on mean Hurst exponent. Bottom right: Time average of grid's frequency spread ($\mathit{TAFD}$, see \eqref{eq:tafd}, normalized to average mean deviation) as coloring.}
\label{fig:micro_net_frq_disp}
\end{minipage}
\end{figure*}

The stability measures typically used in power grid synchronization analysis are not applicable to our stochastic system \cite{hellmann2015survivability,menck2013basin,
nishikawa2006synchronization,pecora1998master,belykh2004connection,Auer2016}. Instead, we use the {\it exceedance} as our main stochastic stability measure to quantify the stability of the synchronous state. It is the cumulated time an observable stays outside a defined ``safe'' region. For our case we define a frequency threshold of $0.01$Hz. This threshold corresponds to the so-called dead band from the German transmission code which defines at which frequency primary control actions kick in to balance deviations from the desired $50$Hz set point \cite{transmission_code}.
As we apply single-node fluctuations for each run, $i=1,..,N$, and record the frequency response for each node $j=1,..,N$, we end up with $N\times N$ frequency time series from which stability measures are derived. In the colorplot of Fig.\ref{fig:stab_measure}(left) the single node exceedance, $E_{i,j}$, is plotted. One grid value represents the probability of network node $j$ to be outside the given frequency band when node $i$ is perturbed: 
\begin{equation}
E_{i,j}=P_i(|f_j|>0.01).
\label{eq:exc}
\end{equation}
This can be further aggregated into the following nodal measures: 
\begin{itemize}
\item The average exceedance over all $N$ nodes given a perturbation at $i$, which we call {\it Troublemaker Index ($\TI$)}: 
\begin{equation}
\bar{E}_i=\frac{1}{N}\sum_{j=1}^N E_{i,j}
\label{eq:ti}
\end{equation}
Power fluctuations at a node with a high $\TI$ causes large frequency deviations often and/or at many nodes. 
\item {\it Excitability} quantifies how much a single-node is exceeding the frequency threshold on average when a random node in the network is perturbed:
\begin{equation}
\bar{E}_j=\frac{1}{N}\sum_{i=1}^N E_{i,j}
\label{eq:excitability}
\end{equation}
Nodes with high excitability react strongly for many origins of the perturbation within the network. We call such nodes highly sensitive.
\end{itemize}
Further, we study the {\it time average of frequency dispersion ($\mathit{TAFD}$)}, the spread in frequency values among between the network nodes averaged over the simulation time
\begin{equation}
\mathit{TAFD}=\frac{1}{T}\sum_{t=0}^{T} \left[\frac{1}{N}\sum_i^N (\Delta f_{i,t}-\mu_t)^2\right]
\label{eq:tafd}
\end{equation} 
where $\Delta f_{i,t}$ is the $i$th node's frequency deviation and $\mu_t$ is the mean frequency deviation, averaged over all nodes, at time $t$. This is a direct measure of the inhomogeneity introduced by the localized fluctuations.

\section{Results}\label{sec:results}
In one simulation run single-node fluctuations are introduced for one specific node. For better comparability the same fluctuation time series is used for each node. Then, the reaction of the whole network towards such fluctuations at the certain node position in the network is investigated. As a result, we find a remarkable interplay of the network structure and the position of the node at which the
fluctuations are fed in. Note that the importance of the network position of nodes, at which power infeed is fluctuating, is an artifact of losses in distribution grids. Fig.\ref{fig:stab_measure} (right) shows a highly homogeneous exceedance for all nodes in lossless grids. Thus, a model setup without lossy lines would miss on the following insights. 

We find network branches with different stability behavior. The top row of Fig. \ref{fig:micro_net_frq_disp} shows branches of troublemakers (with relatively high $\TI$) and branches of vulnerable nodes (with relatively high excitability). The bottom row of Fig. \ref{fig:micro_net_frq_disp} additionally tells us that $\TI$ and the mean Hurst exponent are strongly correlated. Thus, nodes causing  frequency fluctuations above threshold in the whole grid are exactly those that are able to maintain temporal correlations in the frequency time series, $\Delta f_i(t)$, over all grid nodes $i$. 

\begin{figure*}[t!]
\includegraphics[width=0.8\textwidth]{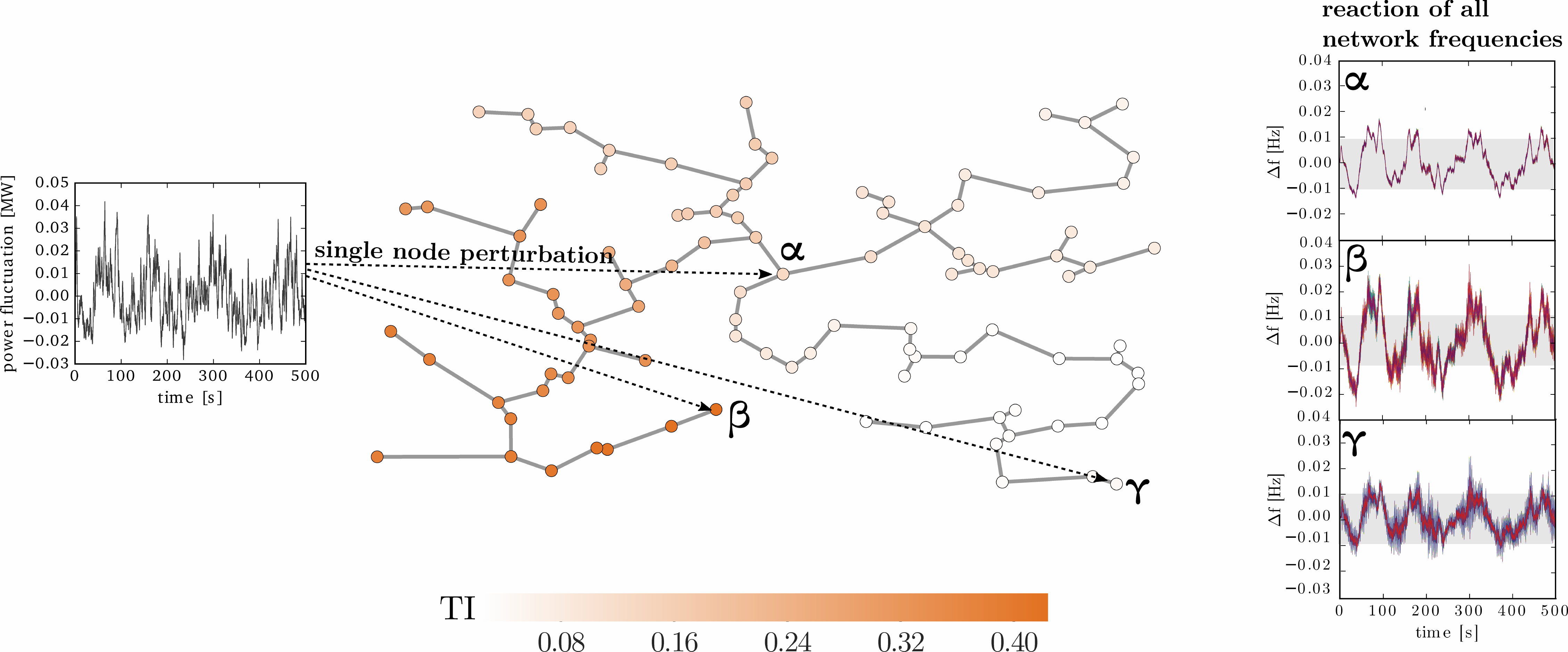}
\caption{Left: Power fluctuation time series $\Delta P(t)$ jointly generated by solar and wind models that capture their intermittent behaviour \cite{anvari2016short,schmietendorf2016}. Center: Random microgrid with $\TI$ as colouring. One simulation run with single node fluctuations at one specific node produces this node's $\TI$ value  Right: Frequency time series for all network nodes $T=500$s for single-node fluctuations at node $\alpha$, $\beta$ and $\gamma$. The grey zone is the frequency threshold band of $0.01$Hz.}
\label{fig:micro_net_exc}
\end{figure*}

At the same time, nodes reacting strongly to intermittent power infeed, $\Delta P(t)$, at whatever node in the grid, themselves, show little ability to be drivers of exceedance. Instead, fluctuations at such nodes lead to large frequency incoherencies or frequency spread ($\mathit{TAFD}$) among the nodes. Hence, high TI does not necessarily mean high frequency spread and vice versa. Nodes, that destabilize the grid, causing large fluctuations at all nodes, are not the same nodes that make the grid incoherent. On the other hand, high $\TI$ nodes are the same nodes that pass on temporal correlations to the other grid nodes.

After these first insights, we want to develop a better understanding of how network structure and high values of $\TI$ and excitability are related in Sections \ref{sec:TI} and \ref{sec:excitability}, respectively.

\begin{figure}[b]
\centering
\hspace{-0.025\textwidth}
\includegraphics[width=0.24\textwidth]{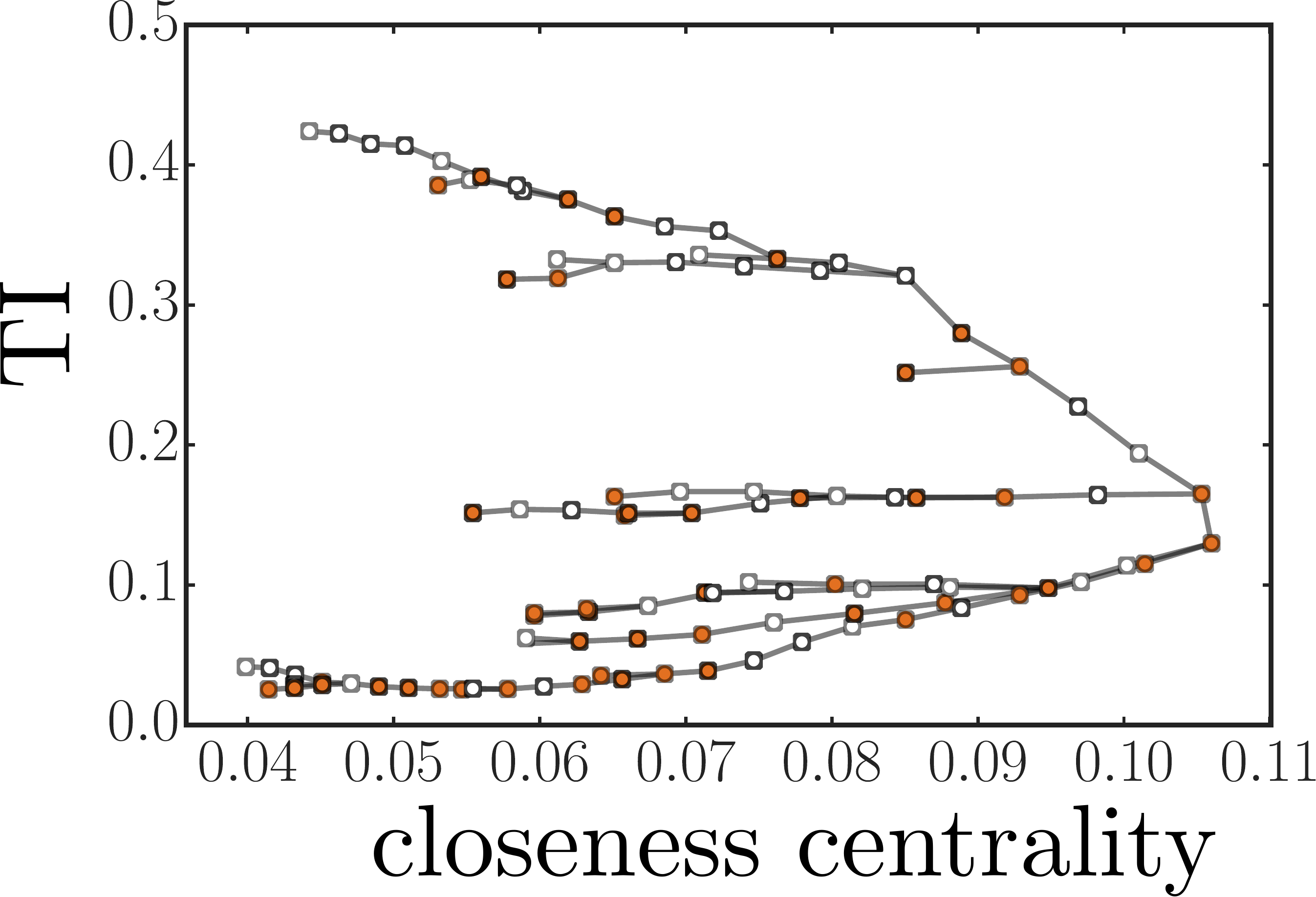}
\hspace{-0.0\textwidth}
\includegraphics[width=0.24\textwidth]{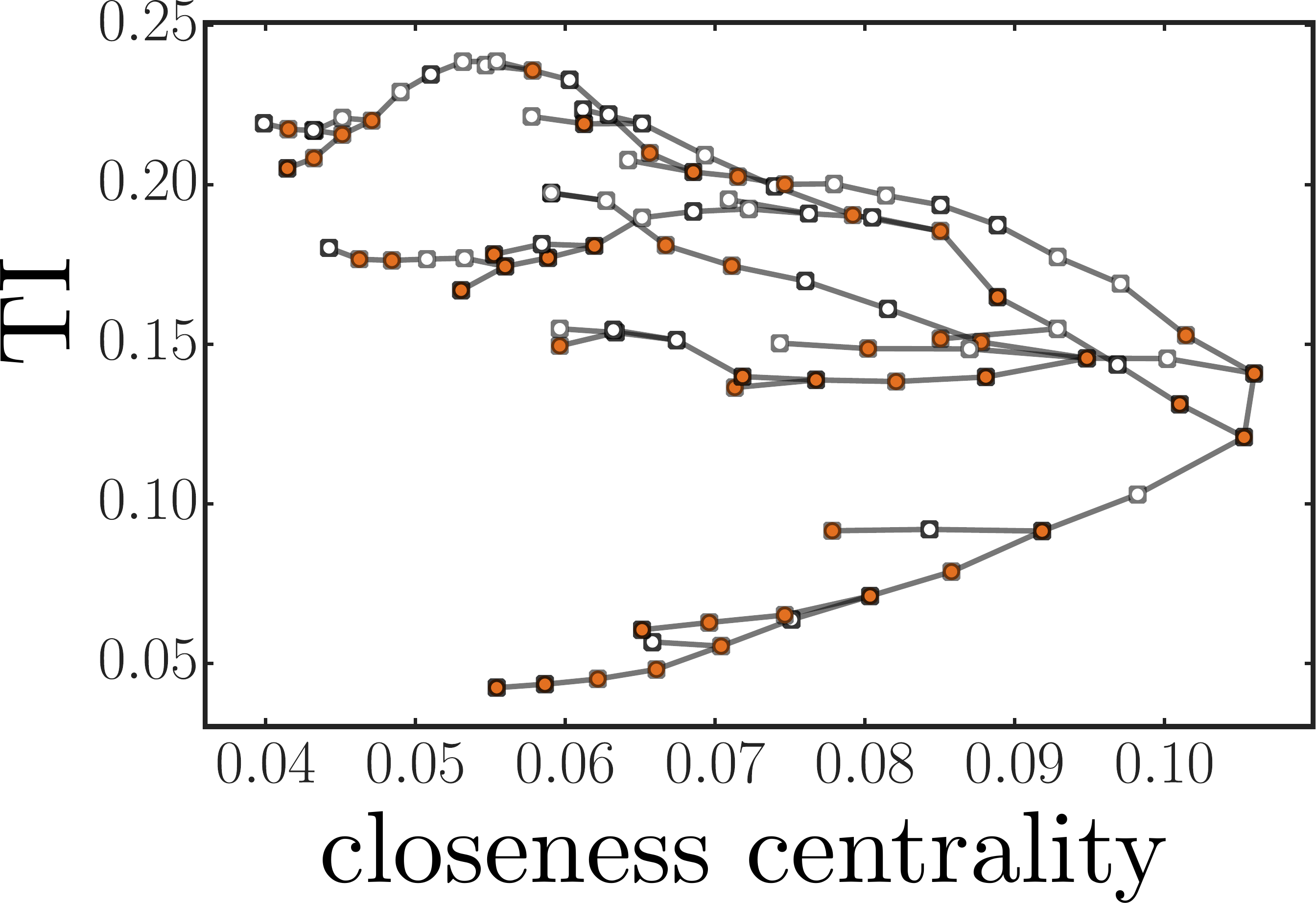}
\hspace{-0.02\textwidth}
\caption{$\TI$ over closeness centrality, $\cc$, for coupling strength for different power input configuration (different realizations of the distribution of consumers and producers) on the same distribution grid. White and orange node represent consumers and producers, respectively.}
\label{fig:net_exc_closeness_diff_config}
\end{figure}

\subsection{The curious tale of trouble-makers}
\label{sec:TI}
\begin{figure}[b]
\centering
\includegraphics[width=0.22\textwidth]{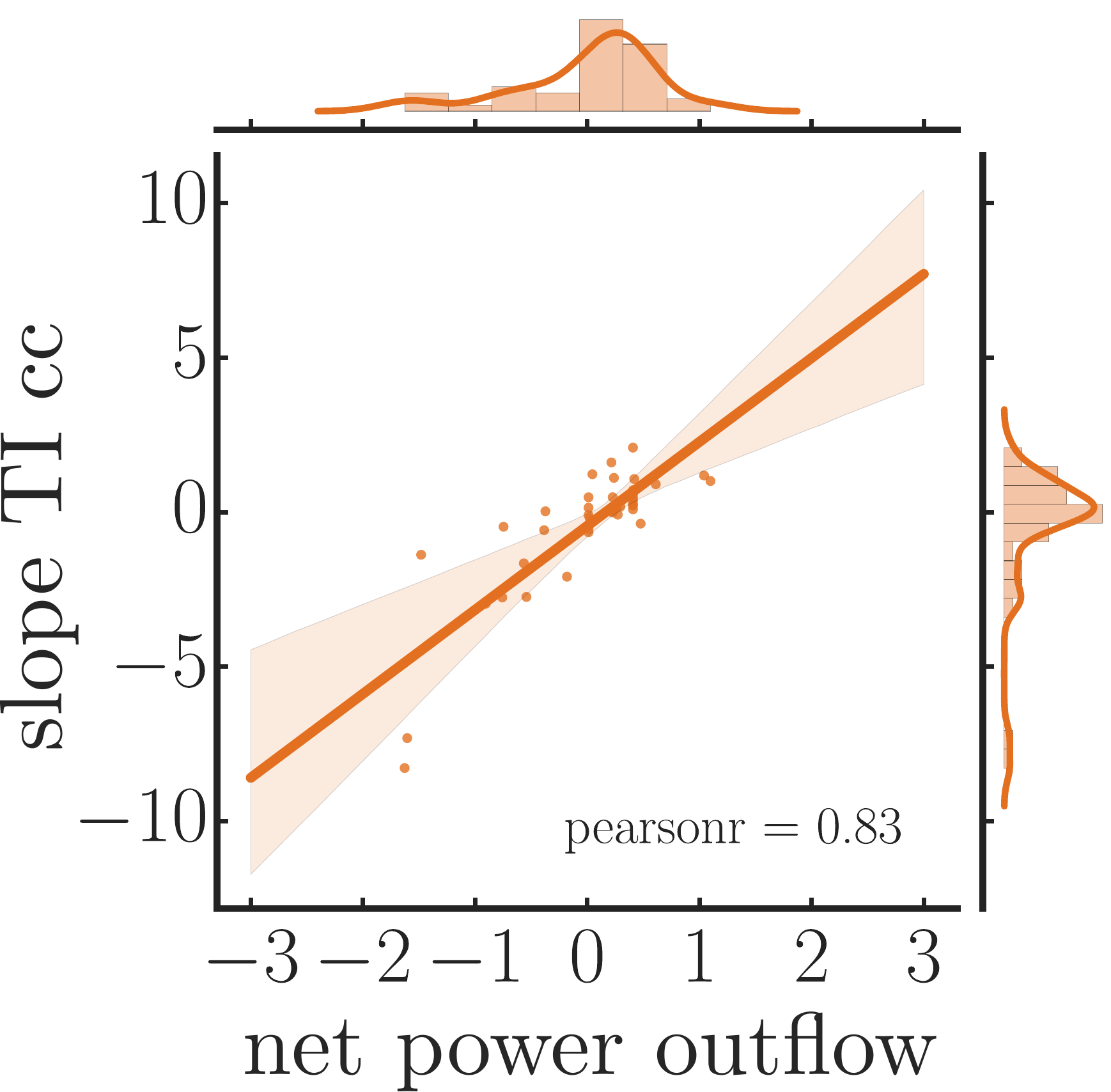}
\hspace{0.02\textwidth}
\includegraphics[width=0.227\textwidth]
{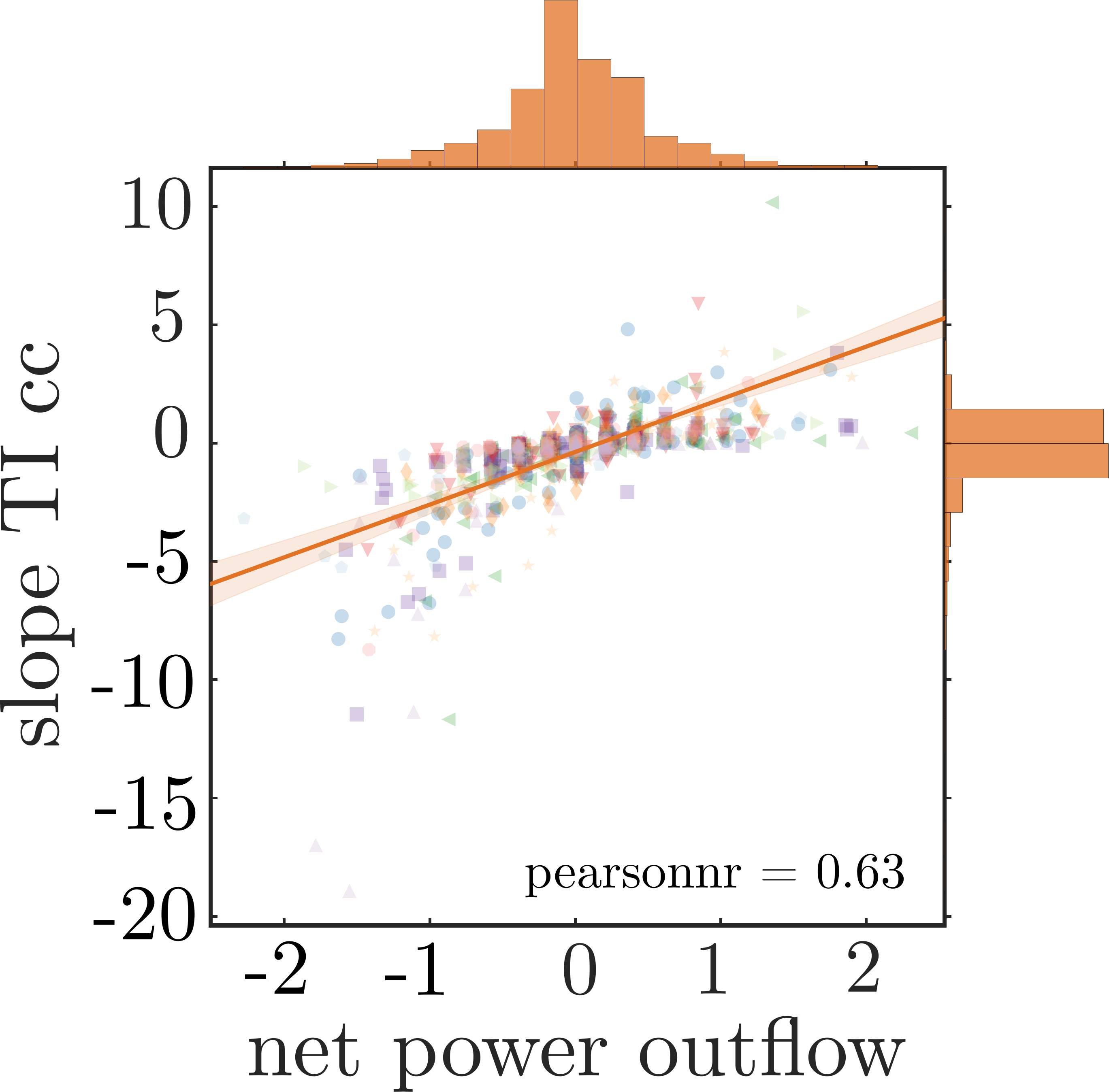}
\caption{Slope of $\TI$ over closeness centrality, $\cc$, (see \eqref{eq:slope}) on the x-axis and branch net power outflow, $P_{out,i}$, (see \eqref{eq:net_power_outflow}) for different distribution grid topologies.}
\label{fig:networks_slope_exc_branch_energy}
\end{figure}
From the time series plots in Fig.\ref{fig:micro_net_exc} (right) it can be seen that it makes a difference in the evolution of frequency deviation on what node the power fluctuations are exerted. The distribution of $\TI$ over the network (Fig. \ref{fig:micro_net_exc} center) shows how nodes of the same branch behave coherently, they have similar capabilities to be drivers of instability.  The frequency time series $\Delta f(t)$ of all network nodes further illustrates the different reaction of the network towards power fluctuations $\Delta P(t)$ at single nodes (see Fig. \ref{fig:micro_net_exc} right). Here, two dead-end nodes from different branches (nodes $\beta$ and $\gamma$) and a node connecting branches, the most closeness central node $\alpha$, were chosen. We see that node $\beta$ is a troublemaker by exceeding the frequency threshold most. $\alpha$ and $\gamma$ exceed less, however, with quite different $\mathit{TAFD}$ values.
The emergence of $\TI$ values different for each branch in the network is remarkable because even for a microgrid with homogeneous power distribution the network topology seems to play an important role.

\begin{figure*}[t!]
\hspace{0.01\textwidth}
\includegraphics[width=0.32\textwidth]{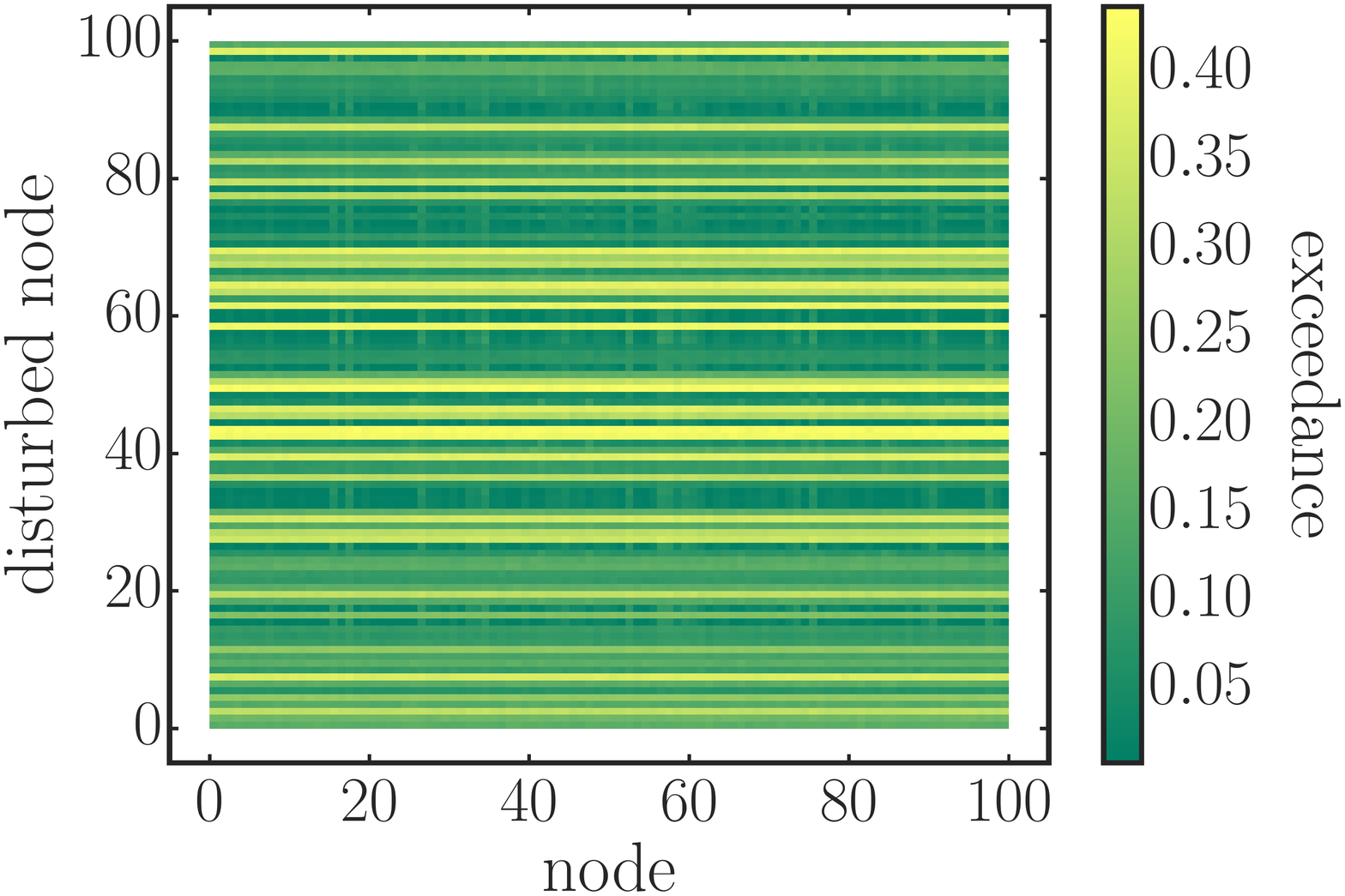}
\includegraphics[width=0.32\textwidth]{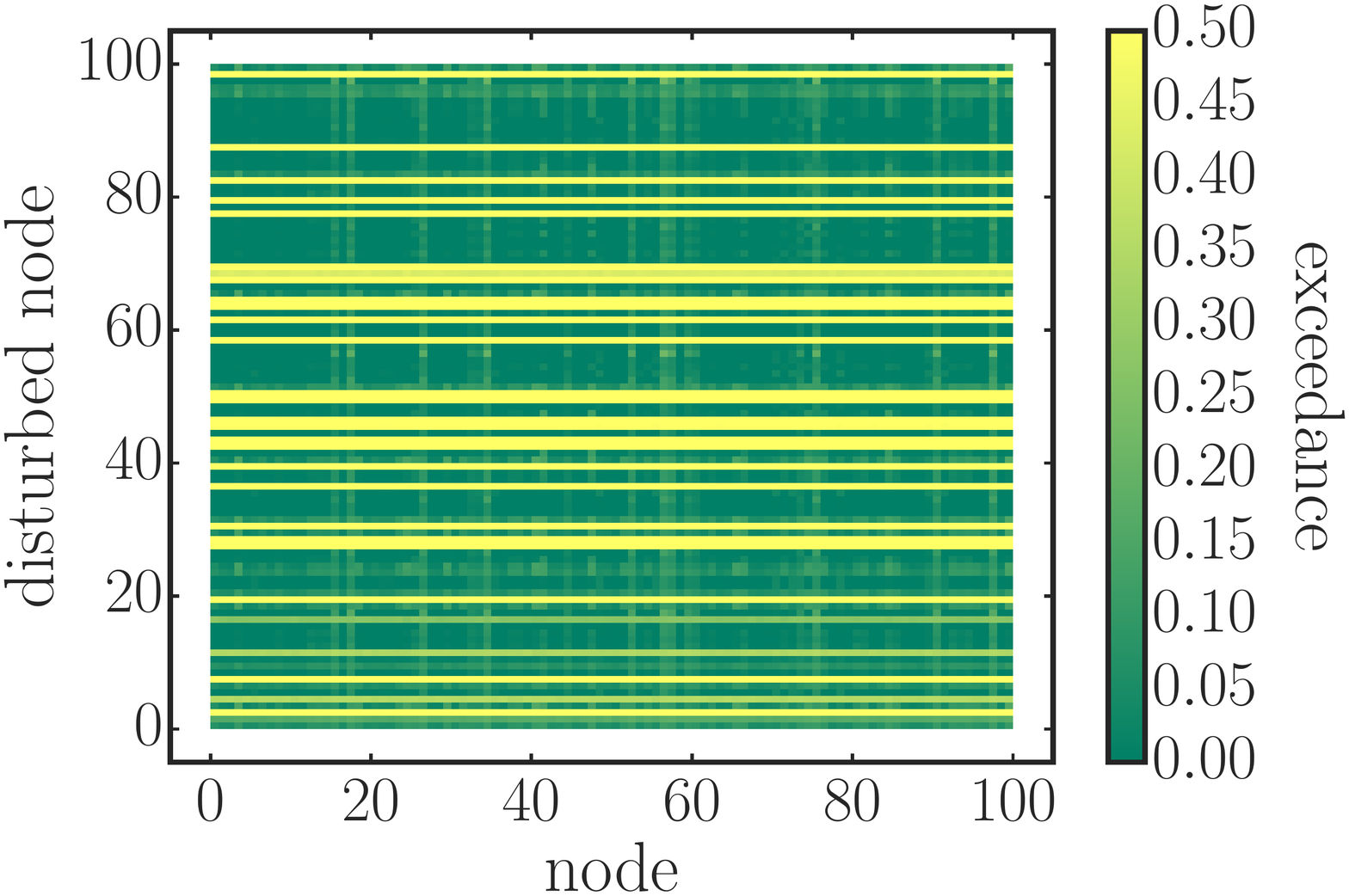}
\includegraphics[width=0.32\textwidth]{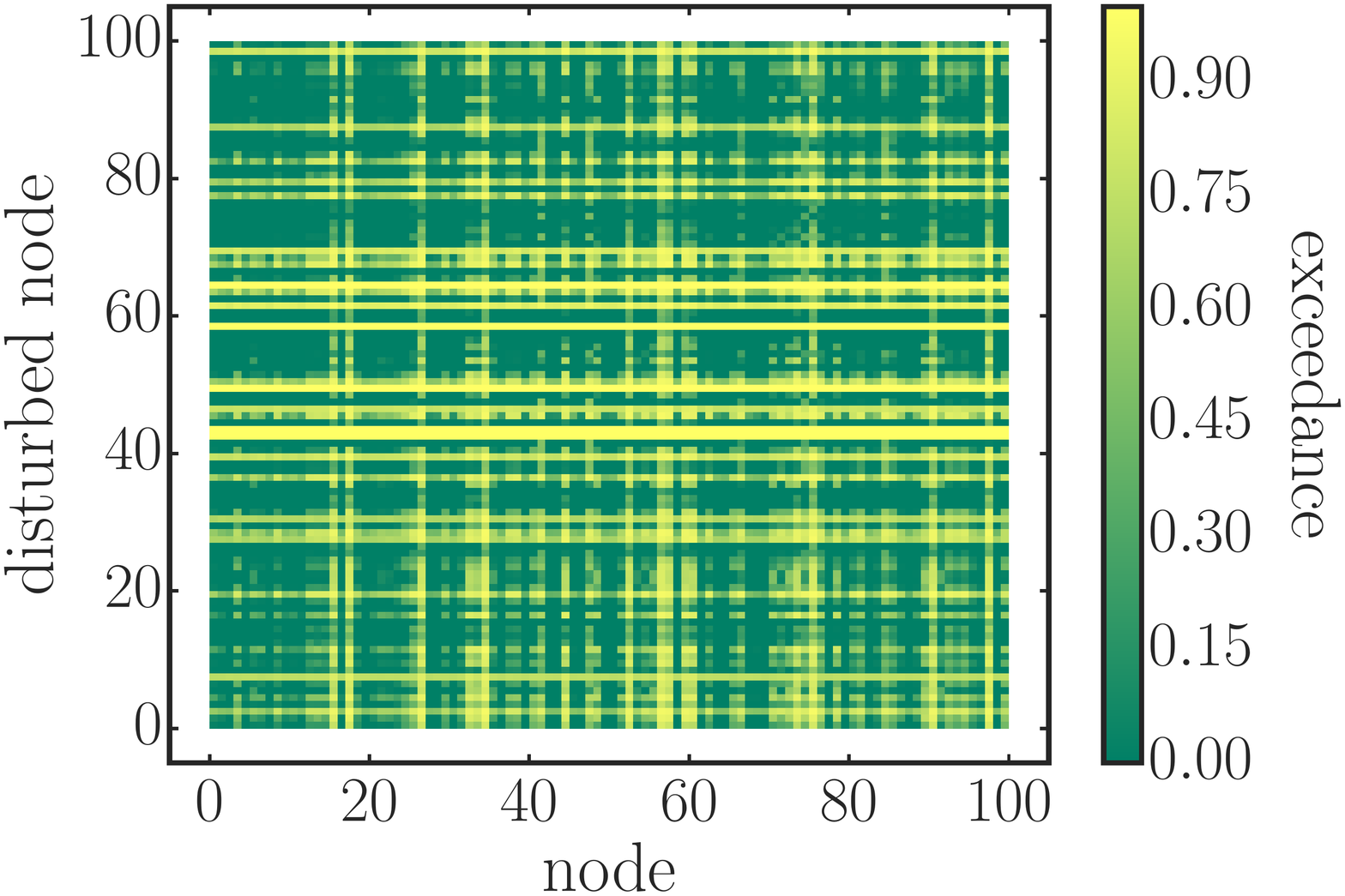}\par
\centering
\includegraphics[width=0.32\textwidth]{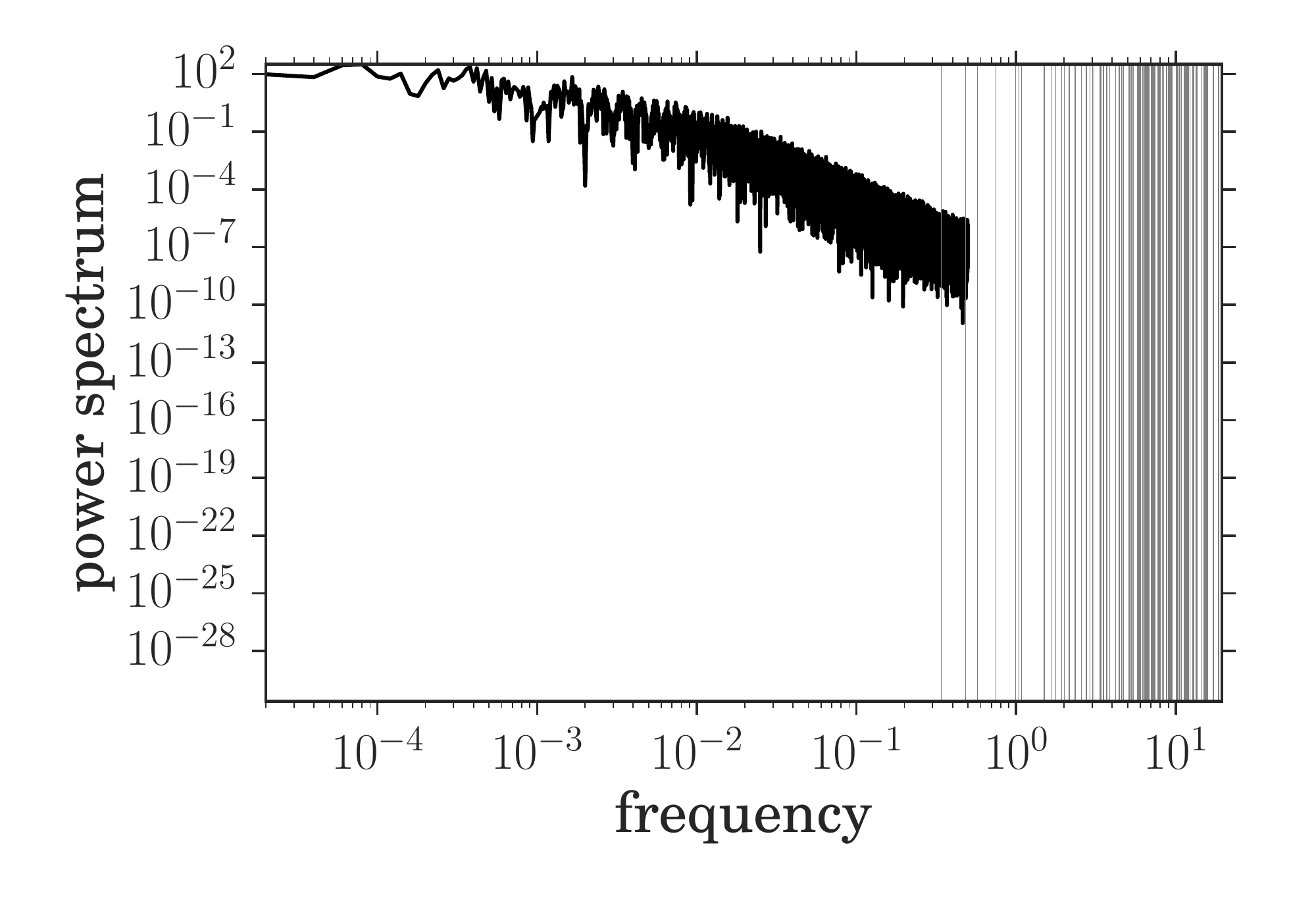}
\includegraphics[width=0.32\textwidth]{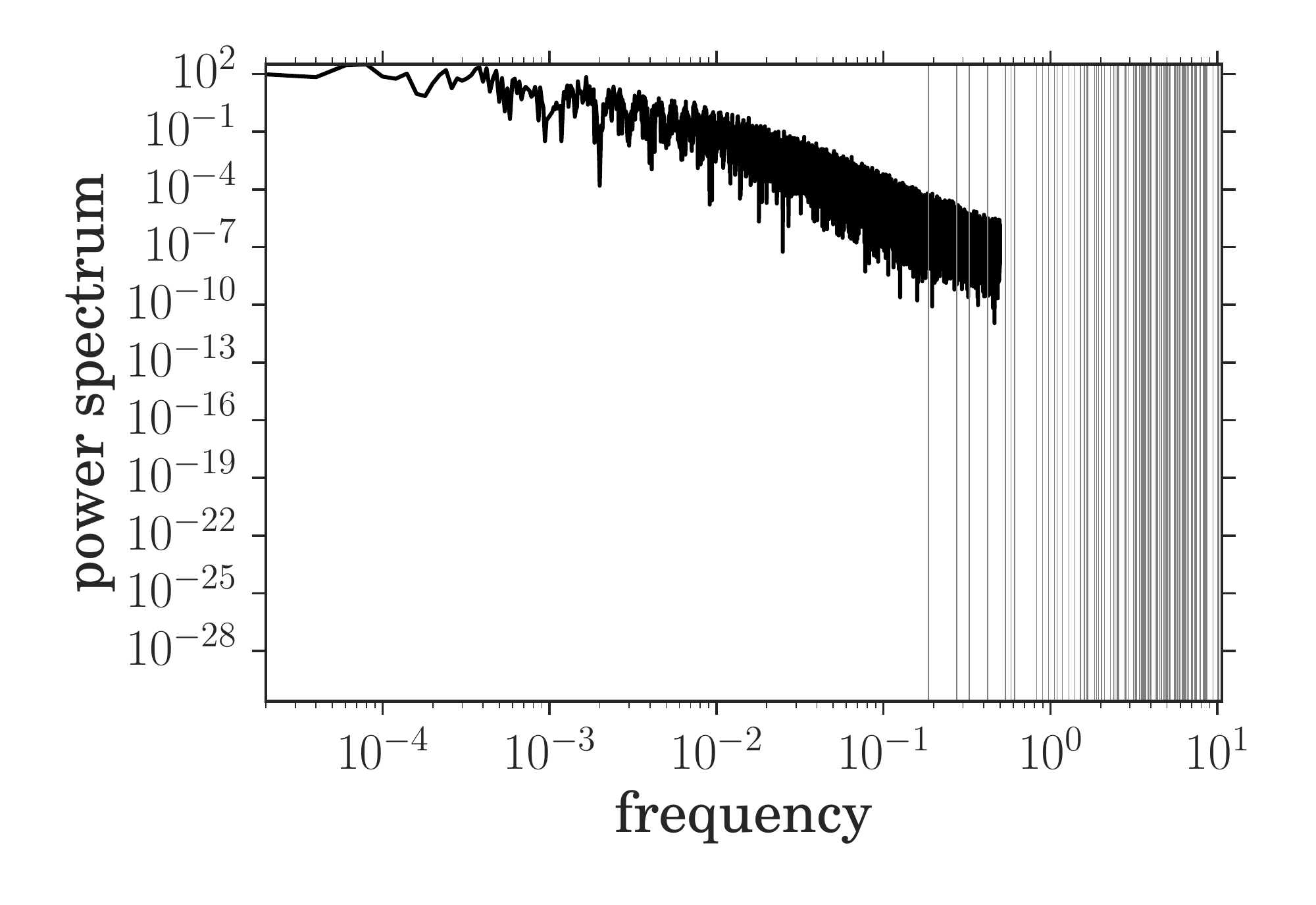}
\includegraphics[width=0.32\textwidth]{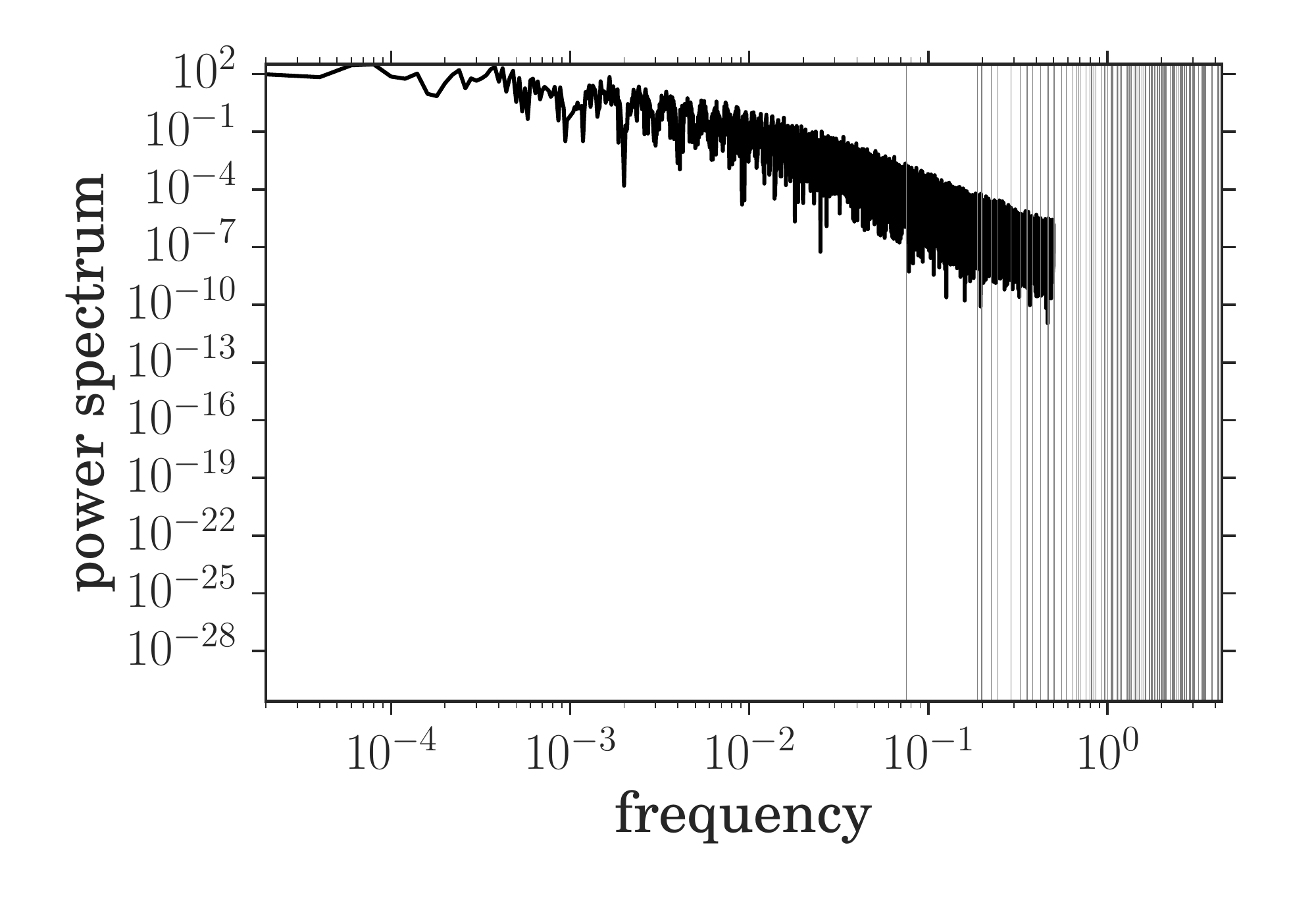}
\caption{Top row: Color plot, with disturbed node on the y-axis and reacting nodes of the network on the x-axis, of single node exceedance. Bottom row: power spectrum of power fluctuations (black) with network eigenfrequencies (grey vertical lines). From left to right: coupling strength in \eqref{eq:dynmics} scaled with factor 1.0, 0.3, and 0.1.}
\label{fig:color_exc_diffcoupling}
\end{figure*}

From Figure \ref{fig:net_exc_closeness_diff_config} (top left) a clear but non-trivial relationship between $\TI$ and closeness centrality is visible. The closeness centrality, $\cc_i$, of node $i$ is defined as the inverse sum over all shortest paths between node $i$ and all other nodes $j$ of the network \cite{newman2003structure}. Therefore, a large $\cc$ value characterizes a node with short distances to all other nodes of the network. Connecting nodes adjacent to each other in the $\TI(\cc)$-plot gives additional information (see Fig.\ref{fig:color_exc_diffcoupling}) . It underlines how branches of high $\TI$ are actual physical network branches. The highest centrality node separates the network into branches of troublemaker nodes and low $\TI$ branches. Hence, certain but not all nodes lead to a temporary desynchronization, even if the frequency is mostly in the bulk regime (see Fig. \ref{fig:color_exc_diffcoupling}). However, why network branches are shifted by different constant factors in $\TI$ is not clear at first sight and is investigated in the following.

From Fig.\ref{fig:net_exc_closeness_diff_config} it becomes evident that different power input configurations play a central role in this phenomenon because a change in the power input configuration alters the branches' $\TI$ values.
The appearance of consumers or producers along the lines towards its dead ends seems to increase or decrease $\TI$, respectively. Due to the losses that are compensated by an equal extra production by all nodes, there is a small asymmetry between consumer and producer power inputs.

Nevertheless, Fig.\ref{fig:net_exc_closeness_diff_config} shows how branches split up with varying slopes for decreasing closeness centrality ($\cc$) .  Hence, a relationship between $\TI$-$\cc$-slope and branch net power outflow can be established. In Fig.\ref{fig:networks_slope_exc_branch_energy} (left) this relationship is plotted for one example grid. The $\TI$-$\cc$-slope is calculated for branch parts as
\begin{equation}
\mathit{slope}_{i,j} =  \frac{\TI_j-\TI_i}{\cc_j-\cc_i}
\label{eq:slope}
\end{equation}
 which lie between two node pairs $(i,j)$ with degree $k\geq 3$ because starting from the most closeness central node each branching seems to alter the slope. This slope is then compared with the net power outflow 
\begin{equation}
P_{out,i}=\sum_{k \ \in \ B_i} P_k
\label{eq:net_power_outflow}
\end{equation}
or the sum over all power inputs of all nodes between the branch part node $i$ with highest $\cc$ in the branch and all other branch nodes, $k$, which are elements of the node set $B_i$ with $\cc_k<cc_i$. Fig.\ref{fig:networks_slope_exc_branch_energy} shows how small slopes are well correlated with net power outflow but we also find especially large slopes which belonging to branch parts closer to the most closest central node. Nevertheless, the Pearson correlation coefficients is 0.83 for the example grid and on average equals 0.73 for an ensemble of 30 distribution grids. Also, the rule is always: branches with negative net power outflow above small values always have negative slopes and vice versa.

\subsection{High excitability and Network Eigenmodes}
\label{sec:excitability}

\begin{figure}[b!]
\begin{minipage}{0.25\textwidth}
\includegraphics[width=0.7\textwidth]{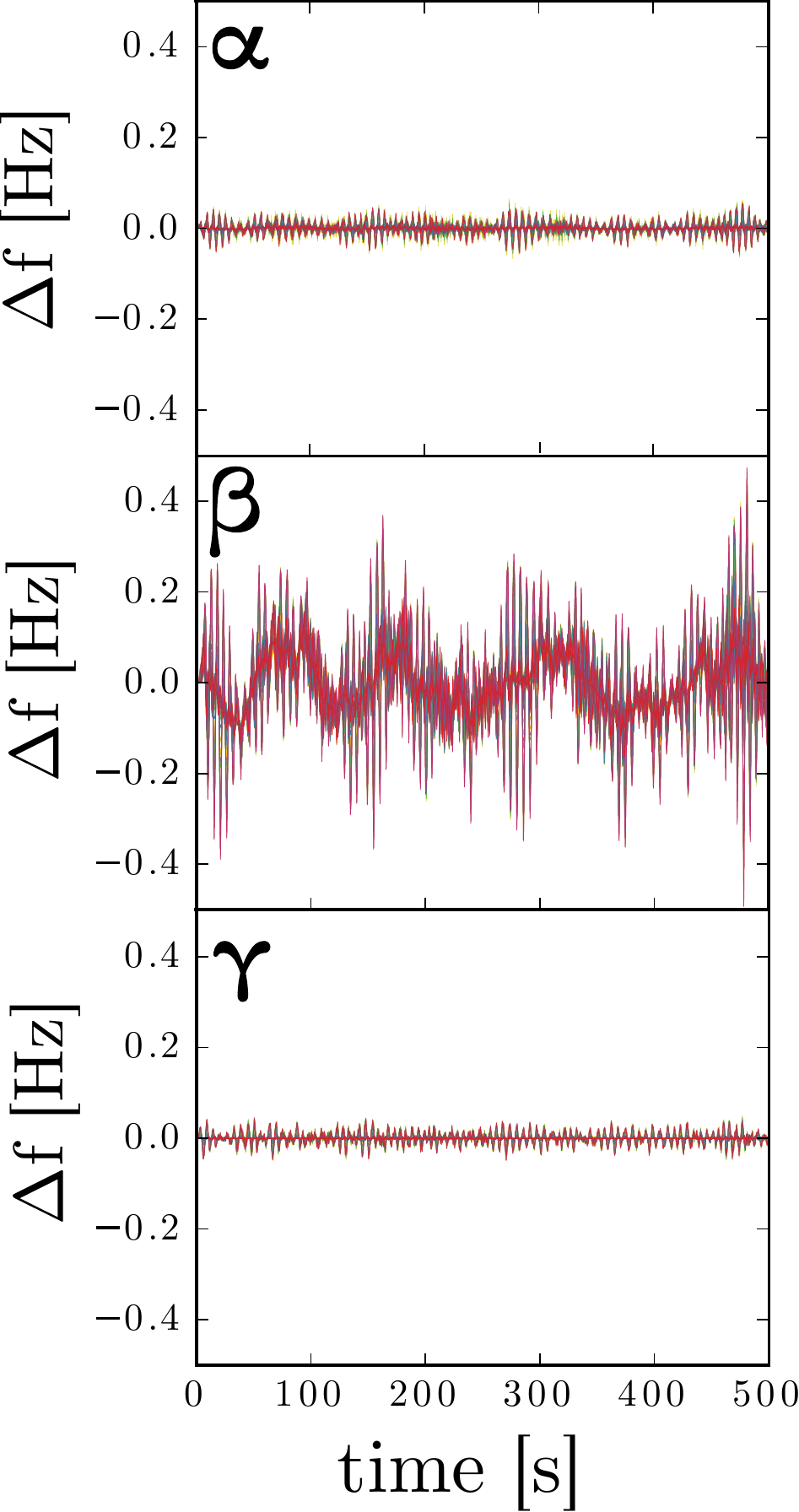}
\end{minipage}
\begin{minipage}{0.2\textwidth}
\caption{Frequency time series for all network nodes for disturbances at node $\alpha$, $\beta$ and $\gamma$. 
with coupling strength reduced to 10\% compared to Fig. \ref{fig:micro_net_exc} of the same distribution grid with identical power input configuration.}
\label{fig:low_coupling_ts}
\end{minipage}
\end{figure}

At first sight, equally reducing the coupling strength leads to much more difference for exerting single-node perturbations to different nodes concerning the network frequency time series (see Fig.\ref{fig:low_coupling_ts}). For comparison, the same nodes as in Fig.\ref{fig:micro_net_exc} were chosen. Compared to node $\alpha$ and $\gamma$, the frequency fluctuations of node $\beta$ are enormously high. For sure, node $\beta$ is a troublemaker because fluctuations at this node lead to all network nodes to be on average 90\% outside the given frequency band ($\TI$ value of 0.9). However, this node has been known as a troublemaker before, for higher coupling strength.

The color plots of Fig. \ref{fig:color_exc_diffcoupling} illustrate how the reduction of coupling strength also brings another aspect into play. For high coupling strength it is clearly visible how some nodes are the driver of high single node exceedance values for all nodes in the network. Now, lowering coupling strength results in two node classes. For low coupling strength there are drivers of instability and nodes that generally tend to be unstable irrespective of which node was initially perturbed. This is underlined by the excitability network plot of Fig. \ref{fig:micro_net_exc}.

This is particularly interesting taking into account the power spectrum of the power fluctuations from wind and solar generation and comparing these with the eigenfrequencies of the Jacobian (see $J_{ij}$ of \eqref{eq:jac}) of the distribution grid dynamics. 
Lowering coupling strength then shifts more and more eigenfrequencies into the power spectrum range of the power fluctuations (see bottom row of Fig. \ref{fig:color_exc_diffcoupling}). Thus, the fluctuations more and more hit the so-called resonant regime mentioned in \cite{Xiaozhu2017}. 

\begin{figure}[t]
\centering
\hspace{-0.01\textwidth}
\includegraphics[width=0.245\textwidth]{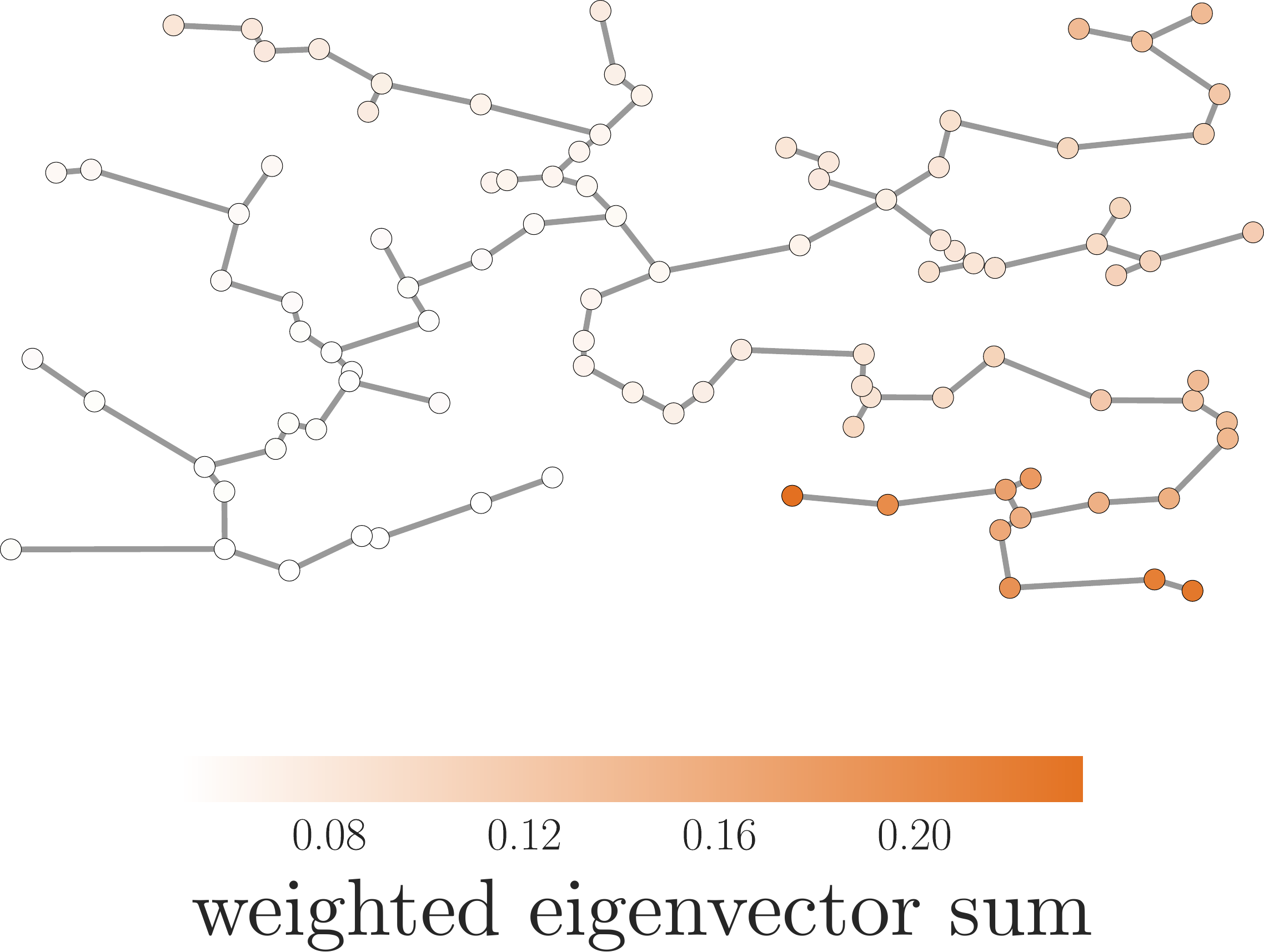}
\hspace{0.01\textwidth}
\includegraphics[width=0.215\textwidth]{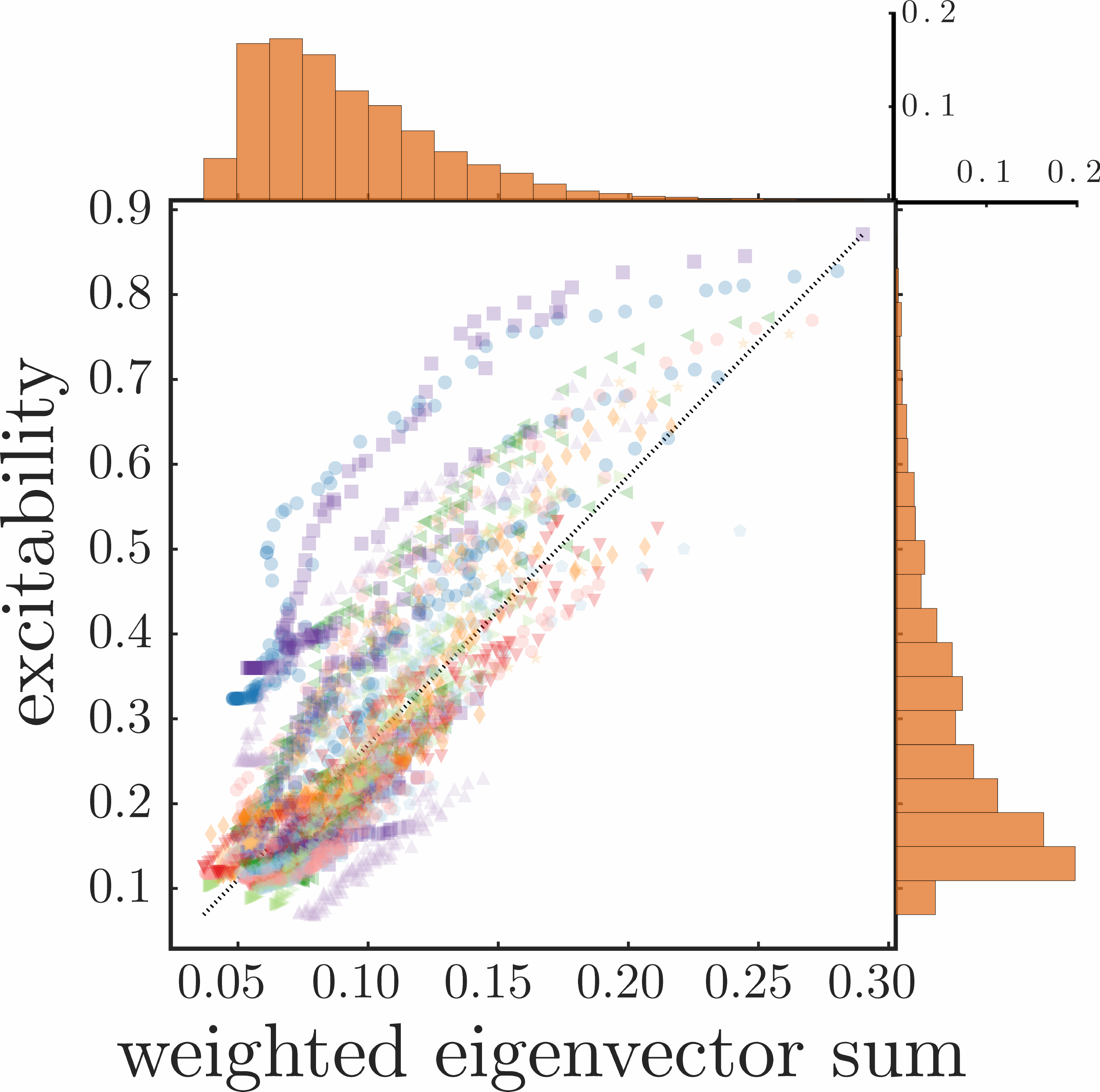}
\caption{Left: The sum of the weighted absolute values of the Jacobian eigenvectors for one example network. Right: Scatter plot of excitability over the weighted Jacobian's eigenvector sum for 30 randomly generated distribution grid trees. All networks have low coupling strength (scaled with factor 0.1) and each color represents one grid.}
\label{fig:network_exc_axis0_disp18}
\end{figure}

The generally instable nodes, appearing as vertical yellow lines in Fig.\ref{fig:color_exc_diffcoupling} (top right), are the nodes with large entries in the Jacobian's eigenvector solutions which provide an analytical predictor for this phenomenon (see Fig.\ref{fig:network_exc_axis0_disp18} (left) for a specific distribution grid tree and Fig.\ref{fig:network_exc_axis0_disp18} (right) for an ensemble of networks). To take into consideration how strong certain eigenmodes of the network can be excited by the power spectrum of the induced fluctuations, a weighted sum (weight is the eigenfrequency to the power of $-5/3$, the Kolmogorov exponent from turbulence theory) of the Jacobian's eigenvectors $\bold{V}_w$ is calculated as:
\begin{equation}
\bold{V}_{w} = \sum_{j=1}^{N}\left[|\Im(\lambda^j)\right|]^{-5/3}|\bold{v}^j|
\end{equation}
where $\Im(\lambda^j)$ is one of the $2N$ Jacobian's eigenfrequency values and $|\bold{v}^j|$ the vector of the magnitudes of the components of the corresponding complex eigenvectors. There are $2N$ eigenvalues but as they appear in complex conjugate pairs we only sum over pairs. 
Concerning the Hurst exponent, it is symmetric towards the perturbed and reacting nodes. This means that correlation in time is preserved for certain nodes irrespective of what node is perturbed initially. The cross correlation between the time series of the disturbed node and all other nodes for different coupling strengths shows how for larger coupling strength cross correlation is generally higher than for networks with lower coupling strength.

\section{Discussion}\label{sec:discussion}

In this work, we showed how important the network position of single-node fluctuations is in terms of its influence on the overall stochastic grid stability. This is a remarkable result for a model case of a microgrid with a homogeneous distribution of consumers and producers. Yet, without the correct representation of distribution grids as lossy networks, this effect would have stayed undiscovered. 

Drivers of instability, the so-called troublemakers and fluctuation sensitive nodes appear on branches, which demonstrated coherent behavior within themselves. The most closeness central node has proven to play a special in role in this type of investigation because it splits the network in branches of such different behavior. The nature of troublemakers is based on a combination of the net power outflow of their corresponding branch part combined with their network centrality. Thus, different power input configuration lead to different results. Also, drivers of instability tend to pass on the temporal correlation of the intermittent power time series to the other grid nodes. 

At the same time, at low coupling strength generally fluctuation sensitive nodes emerge. Such nodes of high excitability themselves have little capability to act as troublemakers. Instead they cause large frequency incoherencies. The reason for their appearance is a strong overlap of the network's eigenfrequencies and the power fluctuation power spectrum for low coupling strength. This leads to large entries in the corresponding Kolmogorov weighted Jacobian's eigenvectors sum that enables us to identify the most excitable or fluctuation sensitive nodes. This gives us an analytical predictor which network regions are especially effected.

In future work, we want to improve our analytic understanding of how the network structure determines troublemakers. Also, a study on multiple node fluctuations will follow which asks for a better understanding of the spatial correlation between fluctuations in renewable energy production. To capture the effect of future voltage issues, as a next step we will introduce voltage dynamics in our power grid simulation. 

From first simulation results, we could see that also meshed grids show the previously described effects, only less pronounced and with reduced values in exceedance. This means, our findings are not restricted to pure tree-like networks. Still, a more detailed investigations of meshed grids, where the identification of branches for evaluation is not possible anymore, shall follow. At the same, such study poses the question how the smart placement of few additional lines may eliminate troublemakers.

There is little known about the frequency dynamics in distribution grids. However, results of our and related work help to reduce balancing needs and improved placement of network stabilizing power balancers and cost-efficient control techniques. 

\section{Acknowledgments}
S.A. wants to thank her fellow colleagues Paul Schultz and Jobst Heitzig for helpful discussion and comments.
The authors gratefully acknowledge the support of BMBF, CoNDyNet, FK. 03SF0472A and the European Regional Development Fund (ERDF), the German Federal Ministry of Education and
Research and the Land Brandenburg for supporting this project by providing resources on the high performance computer system at the Potsdam Institute for Climate Impact Research.

\bibliographystyle{unsrt}
\bibliography{Literature}

\end{document}